\documentclass[prl,aps,twocolumn,superscriptaddress,longbibliography,notitlepage]{revtex4-1}
\pdfoutput=1
\usepackage[colorlinks,linkcolor=darkblue,citecolor=darkblue,urlcolor=darkblue]{hyperref}
\usepackage{amsmath}
\usepackage{xcolor}
\usepackage{verbatim}
\usepackage{graphicx}
\usepackage{esint}
\usepackage{float}
\usepackage{bm}
\usepackage{bbm}
\definecolor{darkblue}{rgb}{0,0,0.6}

\begin{document}

\title{Disordered collective motion in dense assemblies of persistent particles}

\author{Yann-Edwin Keta}

\affiliation{Laboratoire Charles Coulomb (L2C), Université de Montpellier, CNRS, 34095 Montpellier, France}

\author{Robert L. Jack}

\affiliation{Yusuf Hamied Department of Chemistry, University of Cambridge, Lensfield Road, Cambridge CB2 1EW, United Kingdom}

\affiliation{Department of Applied Mathematics and Theoretical Physics, University of Cambridge, Wilberforce Road, Cambridge CB3 0WA, United Kingdom}

\author{Ludovic Berthier}

\affiliation{Laboratoire Charles Coulomb (L2C), Université de Montpellier, CNRS, 34095 Montpellier, France}

\affiliation{Yusuf Hamied Department of Chemistry, University of Cambridge, Lensfield Road, Cambridge CB2 1EW, United Kingdom}

\date{\today}

\begin{abstract}
We explore the emergence of nonequilibrium collective motion in disordered non-thermal active matter when persistent motion and crowding effects compete, using simulations of a two-dimensional model of size polydisperse self-propelled particles. In stark contrast with monodisperse systems, we find that polydispersity stabilizes a homogeneous active liquid at arbitrary large persistence times, characterized by remarkable velocity correlations and irregular turbulent flows. For all persistence values, the active fluid undergoes a nonequilibrium glass transition at large density.  
This is accompanied by collective motion, whose nature evolves from near-equilibrium spatially heterogeneous dynamics at small persistence, to a qualitatively different intermittent dynamics when persistence is large.  This latter regime involves a complex time evolution of the correlated displacement field.
\end{abstract}

\maketitle

Active matter constitutes a prominent class of nonequilibrium systems~\cite{marchetti2013hydrodynamics}.  Examples in biological systems range from tissues~\cite{angelini2011glass} to bird flocks~\cite{cavagna_bird_2014} and human crowds \cite{bain_dynamic_2019}. Synthetic active systems include robot swarms~\cite{rubenstein_programmable_2014} and self-catalytic colloids, which have inspired numerous theoretical studies. In particular, self-propelled particles are paradigmatic models for active matter, where varying a few parameters leads to rich collective behaviour~\cite{fodor_statistical_2018}.

This work focuses on the interplay of motility and crowding in such models, which is relevant in both synthetic~\cite{klongvessa2019active} and living systems~\cite{angelini2011glass}.
For systems with aligning interactions or non-spherical particles, collective motion occurs at large density~\cite{vicsek2012collective,bar2020selfpropelled}, as predicted theoretically~\cite{dunkel2013fluid} and observed in experiments~\cite{wensink2012meso,angelini2011glass} and simulations~\cite{wensink2012emergent,mandal2017glassy}.
For isotropic particles, it is well-known that persistent motility can trigger phase separation (MIPS)~\cite{cates2015motility}; it also leads to nonequilibrium velocity correlations~\cite{szamel2021long}, possibly connected to active turbulence~\cite{mandal2020extreme,kuroda2022anomalous}.
A different set of phenomena are associated with crowding effects in systems with modest persistence times: they undergo nonequilibrium glass transitions at high density~\cite{berthier2019glassy,janssen2019active}.  This transition shares many features with its equilibrium counterpart~\cite{berthier2011theoretical}, such as two-step relaxation dynamics and slow, spatially correlated motion. It has been thoroughly studied using theory~\cite{berthier2013non,nandi_random_2018,reichert2021transport}, simulations~\cite{berthier2014nonequilibrium,flenner2016nonequilibrium,berthier2017active,berthier2019glassy}, and experiments~\cite{klongvessa2019active,klongvessa2019nonmonotonic}.

\begin{figure}[b]
\includegraphics[height=4cm,width=1.0\columnwidth]{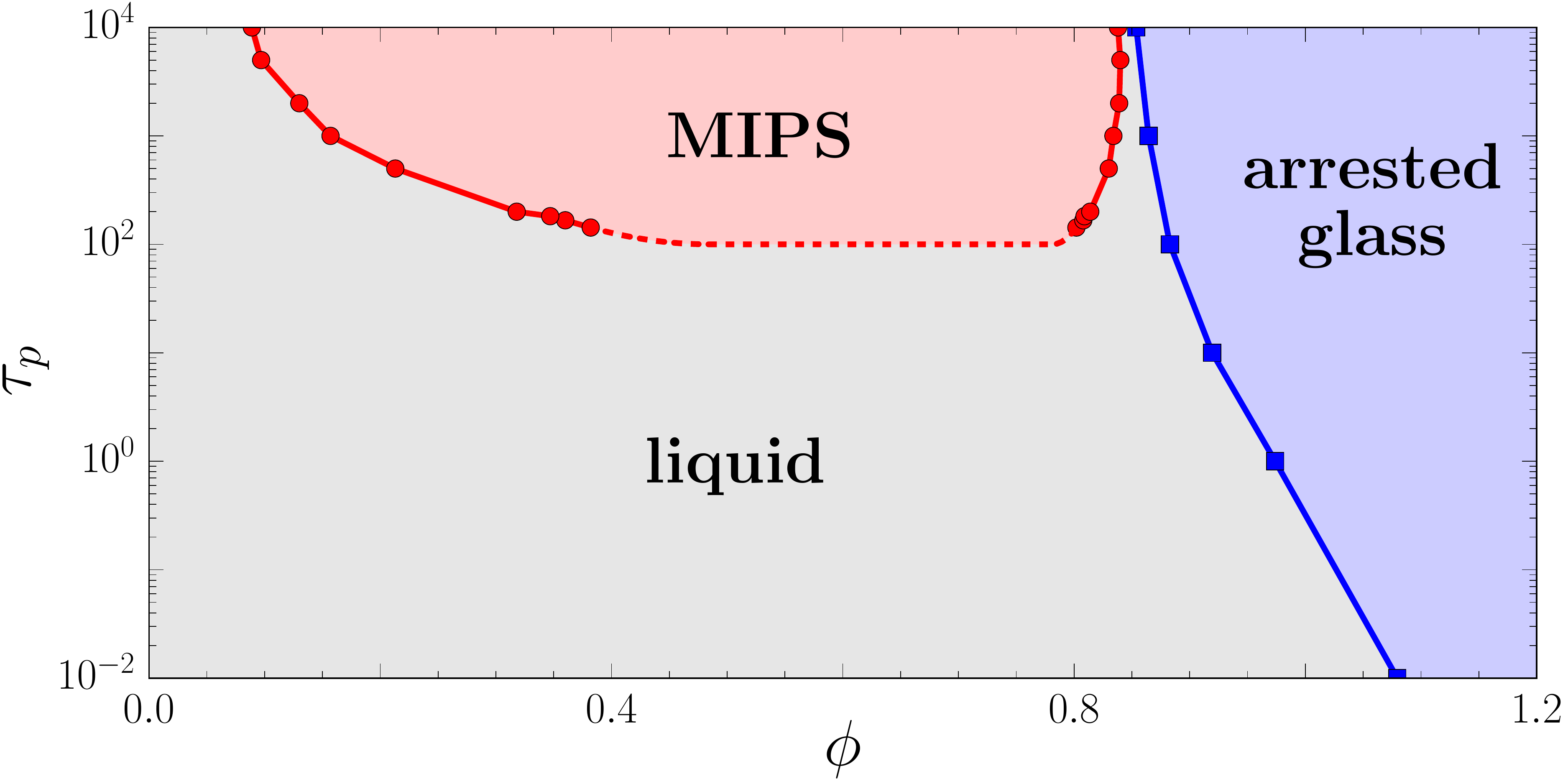}
\caption{Disordered active states when packing fraction $\phi$ and persistence time $\tau_p$ are varied at constant $D_0=1$. The MIPS region, enclosed by red circles, corresponds to phase separation. The homogeneous active liquid experiences a nonequilibrium glass transition at large packing fraction (blue squares).}
\label{fig:pd}
\end{figure}

In the following we present numerical simulations of isotropic particles that unify this active glass phenomenology with the effects of crowding at much larger persistence times, where recent exploratory works have suggested the emergence of intermittent plasticity~\cite{mandal2020extreme}, analogies to yielding transitions~\cite{liao2018criticality,morse2021direct,villarroel2021critical}, and a breakdown of equipartition in the disordered solid~\cite{henkes2020dense,caprini2020active}. We intentionally introduce size polydispersity, to ensure the stability of disordered states in crowded systems.  Hence, our results (\textit{e.g.} Fig.~\ref{fig:pd}) parallel the similar programme that was recently completed for monodisperse particles, which display ordered phases at large density~\cite{redner2013structure,digregorio2018full,omar_phase_2021,caprini2020hidden,caprini2020spontaneous}.

\begin{figure*}
\centering
\includegraphics{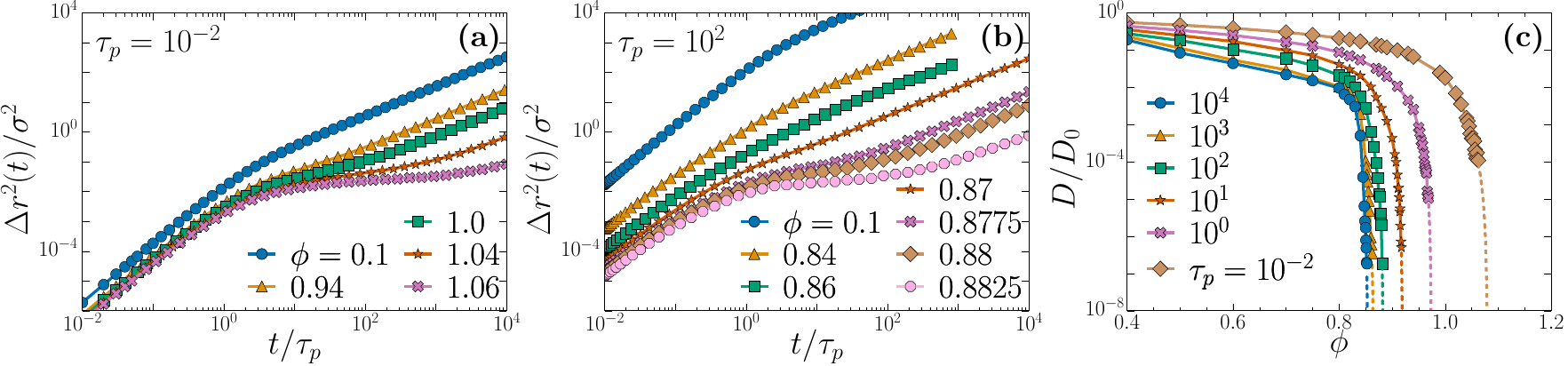}
\caption{(a, b) Mean squared displacements for different packing fractions $\phi$, at persistence time (a) $\tau_p = 10^{-2}$ and (b) $\tau_p = 10^2$, for $N = 1024$. (c) Density evolution of the scaled diffusion constant $D/D_0$ for different persistence times $\tau_p$. Dashed lines correspond to algebraic fits.}
\label{fig:arrest}
\end{figure*}

At large persistence, we find that polydispersity stabilizes a homogeneous liquid ~\cite{fily2014freezing,paoluzzi_are_2021} which solidifies into a disordered glass at large packing fraction (see Fig.~\ref{fig:pd}). This liquid displays inherently active characteristics, including correlated velocities and displacements, with similarities to active turbulence.  Close to dynamic arrest, its relaxation is intermittent, characterized by rapid rearrangements between transient mechanical equilibria~\cite{mandal2020multiple,mandal2021how}.
This behavior is distinct from active glasses at lower persistence, which tend to resemble passive glasses~\cite{berthier2019glassy,janssen2019active}.

We simulate $N$ athermal self-propelled particles in a square box of linear size $L$ with periodic boundary conditions, with overdamped dynamics:
\begin{equation}
\xi \dot{\boldsymbol{r}}_i = - \sum_{j \neq i} \nabla_i U(r_{ij}) + \boldsymbol{p}_i,
\label{eq:r}
\end{equation}
 where
$\boldsymbol{r}_i$ is the position of particle $i$,
$r_{ij} = | {\boldsymbol r}_i - {\boldsymbol r}_j|$, $\xi$ a viscous damping, and $\boldsymbol{p}_i$ the propulsion force. Particles interact via a repulsive Weeks-Chandler-Andersen potential $U = 4 \varepsilon \left[ (\sigma_{ij}/r_{ij})^{12} - (\sigma_{ij}/r_{ij})^6 + 1/4 \right]
\Theta(2^{1/6}\sigma_{ij}-r_{ij})$, where $\Theta$ is the Heaviside function, $\sigma_{ij} = \frac{1}{2} (\sigma_i + \sigma_j)$, and $\sigma_i$ the diameter of particle $i$. Diameters are drawn from a uniform distribution of mean $\sigma=\overline{\sigma_i}$ and polydispersity 20\%. The packing fraction is $\phi = 2^{1/3} \pi N \overline{\sigma_i^2}/(4 L^2)$, $\sigma$ sets the unit length, $\varepsilon$ the unit energy, and $\xi \sigma^2 / \varepsilon$ the unit time. We measure the positions $\boldsymbol{r}_i$ and velocities $\boldsymbol{v}_i$ in the center-of-mass frame.

Different self-propulsion models differ by the dynamics of $\boldsymbol{p}_i$. For active Brownian and run-and-tumble particles~\cite{solon_active_2015},  $|\bm{p}_i|$ is fixed.  We consider active Ornstein-Uhlenbeck particles~\cite{szamel_self-propelled_2014,koumakis2014directed}:
\begin{equation}
\tau_p \dot{\boldsymbol{p}}_i = - \boldsymbol{p}_i + \sqrt{2 D_0} \boldsymbol{\eta}_i ,
\label{eq:p}
\end{equation}
with $\boldsymbol{\eta}_i$ a Gaussian white noise of zero mean and variance $\left<\boldsymbol{\eta}_i(t) \boldsymbol{\eta}_j(0)\right> = \mathbbm{1} \, \delta_{ij} \, \delta(t)$, and $\tau_p$ the persistence time.  The amplitude and correlation time of $\boldsymbol{p}_i$ can be varied independently so there are three control parameters $(\phi, D_0, \tau_p)$.
The parameter space reduces to $(\phi, \tau_p)$ for self-propelled hard disks~\cite{berthier2014nonequilibrium}, while adding a thermal noise in Eq.~(\ref{eq:r}) would instead make the phase diagram four-dimensional~\cite{paoluzzi_are_2021}. From Eqs.~(\ref{eq:r}, \ref{eq:p}), we see that $D_0$ is the diffusion constant of a free particle, the typical amplitude of the propulsion is $\sqrt{\left<| {\boldsymbol p}_i|^2\right>} = \sqrt{2 D_0/\tau_p} \equiv v_0$, and the persistence length $\ell_p = \sqrt{D_0 \tau_p}$. When $\tau_p \to 0$, equilibrium Brownian dynamics at temperature $T_{\rm eff} = D_0$ is recovered.  We have numerically explored the full three-dimensional phase diagram of the model and found that its main features are understood by keeping $D_0$ constant and varying the persistence time $\tau_p$. We show results for $D_0=1$, so that equilibrium at $T_{\rm eff}=1$ is recovered when $\tau_p = 0$, but the system is out of equilibrium in the entire  $(\tau_p,\phi)$ phase diagram in Fig.~\ref{fig:pd}.

\begin{figure*}
\centering
\includegraphics{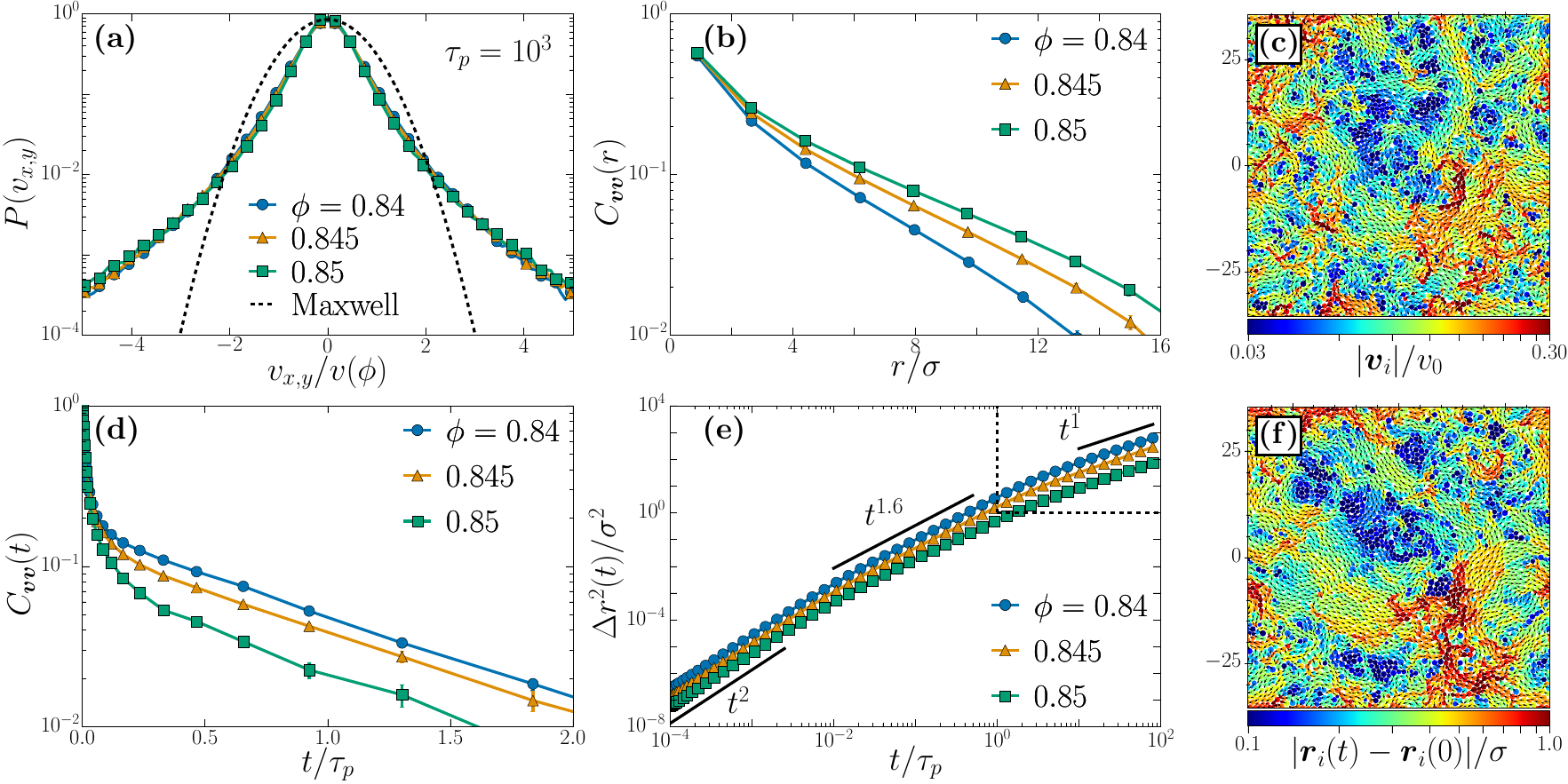}
\caption{
  (a) Velocity distribution $P(v_{x, y})$ for different packing fractions $\phi$ measured along the $x$ and $y$ directions. The dashed line corresponds to the Maxwell distribution.
  (b) Spatial correlations of the velocities $C_{\boldsymbol{v}\boldsymbol{v}}(r) = \left<\boldsymbol{v}_i(0) \cdot \boldsymbol{v}_j(0) \delta(r - r_{ij})\right>/v(\phi)^2 \left<\delta(r - r_{ij})\right>$ with $r_{ij} = |\boldsymbol{r}_i(0) - \boldsymbol{r}_j(0)|$.
  (c) Example of a velocity field for $N = 4096$, $\tau_p = 10^3$, $\phi = 0.84$.
  (d) Time autocorrelations of the velocity $C_{\boldsymbol{v}\boldsymbol{v}}(t) = \left<\boldsymbol{v}_i(t) \cdot \boldsymbol{v}_i(0)\right>/v(\phi)^2$.
(e) Mean-squared displacements for $\tau_p = 10^3$ and $N = 4096$ with ballistic, superdiffusive, and diffusive behaviours indicated.
(f) Displacement field after a time $t = 0.11 \tau_p$ (the MSD is $\Delta r^2(t) = 0.16\sigma^2$) starting from the velocity field shown in (c) at $t=0$.} 
\label{fig:vel}
\end{figure*}

Figure~\ref{fig:pd} shows the three disordered phases of polydisperse self-propelled disks. The homogeneous fluid appears for intermediate packing fraction and persistence time $\tau_p \lesssim 10^2$; it phase separates in the MIPS region at larger persistence. The MIPS boundaries (red circles in Fig.~\ref{fig:pd}) were established by locating the two peaks of the probability distribution function of the local packing fraction~\cite{digregorio2018full}. We only sketch the critical point of MIPS, its precise characterisation requiring a more detailed analysis~\cite{maggi2021universality}. Unlike monodisperse particles, whose relaxation relies on the movement of defects \cite{briand2016crystallization}, the dense phase of MIPS is fully disordered and flows easily: it corresponds to an active liquid phase which exists up to arbitrarily large persistence in between MIPS and dynamic arrest.

For all $\tau_p$, the active liquid undergoes dynamic arrest at large $\phi$ and transforms into an active glass~\cite{berthier2019glassy}. To characterize the liquid-glass boundary, in Fig.~\ref{fig:arrest} we analyze the mean-squared displacement (MSD)
\begin{equation}
\Delta r^2(t) = \frac{1}{N} \sum_i \left<|\boldsymbol{r}_i(t) - \boldsymbol{r}_i(0)|^2\right>,
\label{equ:msd}
\end{equation}
where brackets indicate an ensemble average. In the liquid, the MSD is ballistic at short times, $\Delta r^2 \sim (v t)^2$, and diffusive at long times, $\Delta r^2 \sim 4 D t$. These limits define the mean-squared velocity, $v^2(\phi) \equiv \left<|\boldsymbol{v}_i|^2\right>$, and the self-diffusion constant, $D(\phi)$, which respectively reduce to $v_0^2$ and $D_0$ in the dilute limit, $\phi \to 0$.

In Figs.~\ref{fig:arrest}(a,b), we contrast the $\phi$-dependent MSDs at small and large $\tau_p$. In both cases, $D(\phi)$ decreases by several orders of magnitude when $\phi$ varies over a modest range, see Fig.~\ref{fig:arrest}(c).  As in equilibrium systems, the determination of a critical packing fraction for dynamic arrest is ambiguous~\cite{brambilla_probing_2009}: we fit $D(\phi)$ to an algebraic form inspired by mode-coupling theory, $D \sim (\phi_c - \phi)^{\gamma}$, to estimate a liquid-glass boundary $\phi_c(\tau_p)$ (blue squares in Fig.~\ref{fig:pd}).
The MSD data in Figs.~\ref{fig:arrest}(a,b) both show the emergence at large $\phi$ of a two-step relaxation process separated by a subdiffusive plateau. Despite these similarities, the remainder of this work demonstrates several important differences between the highly-persistent case and weakly persistent or equilibrium glasses.

In equilibrium systems, velocities are spatially uncorrelated and obey the Maxwell distribution with a mean-squared velocity $v^2$ independent of $\phi$.
The situation is very different in persistent systems, where $v^2(\phi)$ decreases sharply with $\phi$, see Fig.~\ref{fig:arrest}(b). This already signals the importance of nonequilibrium effects on the velocities of interacting self-propelled particles~\cite{fodor_statistical_2018,marconi_velocity_2016,caprini2020active}.

In Fig.~\ref{fig:vel}, we analyze particle dynamics in a highly persistent liquid ($\tau_p=10^3$) whose density $\phi$ is between MIPS and dynamic arrest. The velocity distributions in Fig.~\ref{fig:vel}(a) are strongly non-Maxwellian with much broader tails. Since a Gaussian distribution is expected both for interacting equilibrium particles
and for non-interacting Ornstein-Uhlenbeck particles, the measured distributions reveal the non-trivial influence of many-body interactions in persistent liquids \cite{marconi_velocity_2016,caprini2020active}.

Even more interestingly, persistent propulsions in the homogeneous liquid produce spatial velocity correlations~\cite{szamel2015glassy,flenner2016nonequilibrium,henkes2020dense,szamel2021long,caprini2020spontaneous,caprini2020hidden}, which we quantify in  Fig.~\ref{fig:vel}(b) via the velocity correlation $C_{\boldsymbol{v}\boldsymbol{v}}(r)$. The data clearly reveal the existence of correlations extending over a length scale which grows with $\phi$ and $\tau_p$. The corresponding real-space correlations extend over several particle diameters, as illustrated in Fig.~\ref{fig:vel}(c). Such correlated velocity patterns have no equilibrium analog. They can be rationalised via the coupling of collective elastic modes and highly-persistent active forces \cite{henkes2020dense}. Excitation of these modes by thermal fluctuations affects displacements in passive glasses~\cite{shiba2016unveiling,illing2017mermin}: the Supplemental Material (SM) \cite{suppmat} shows that their effects are weak here.

\begin{figure}
\centering
\includegraphics{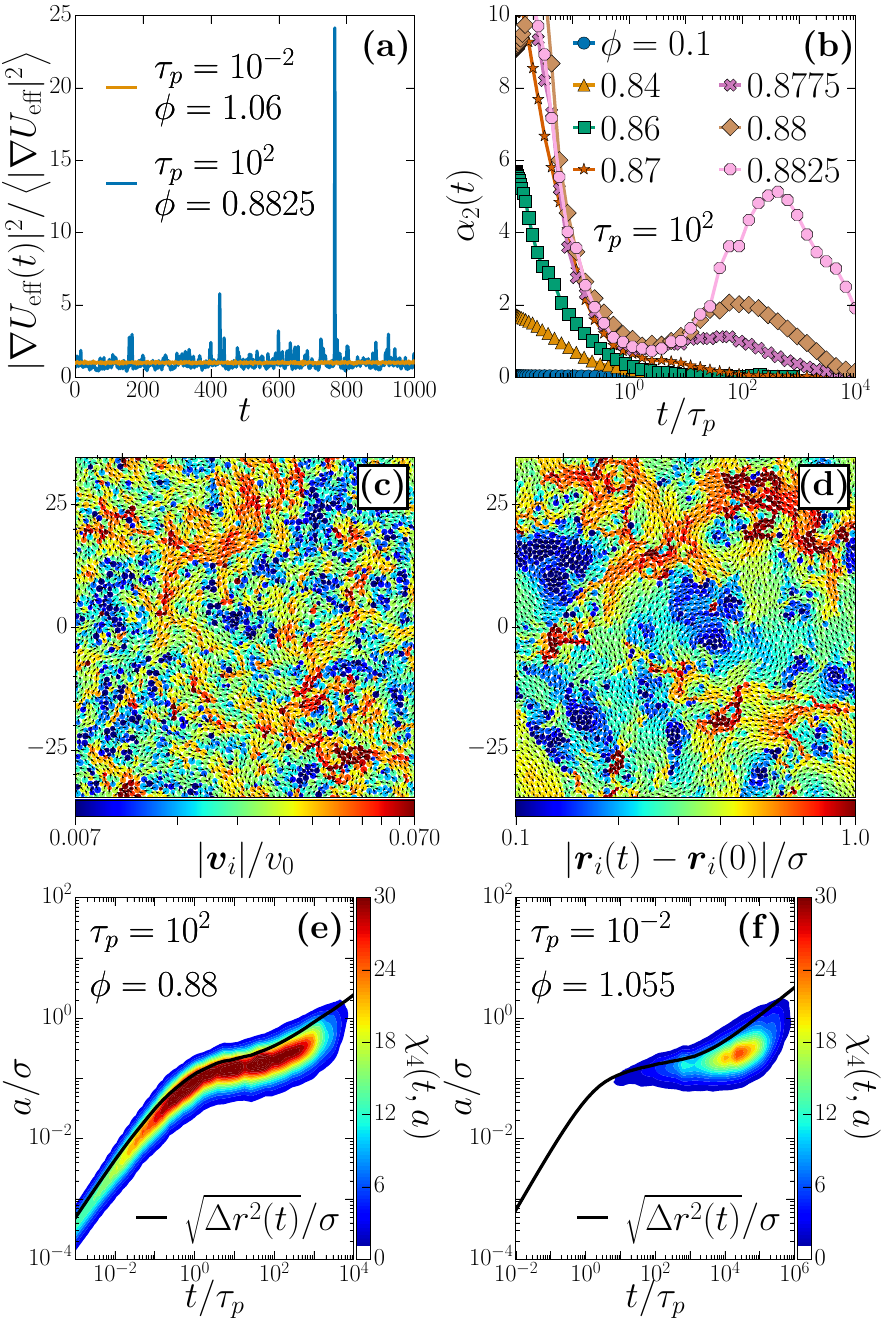}
\caption{(a) Time series of the kinetic energy $\sum_i |\boldsymbol{v}_i(t)|^2 = |\nabla U_{\mathrm{eff}}(t)|^2$ rescaled by its time average, with $N = 1024$ for two dense systems at $\tau_p = 10^{-2}, 10^2$.
(b) Non-Gaussian parameter $\alpha_2(t)$ for $\tau_p = 10^2$, $N = 1024$ and different packing fractions.
(c) Velocity field for $N = 4096$, $\tau_p = 10^2$, $\phi = 0.8825$.
(d) Displacement field after a time $t=200 \tau_p$ (The MSD is $\Delta r^2(t) = 0.13\sigma^2$) starting from the velocity field shown in (a) at $t=0$
(e) Dynamical susceptibility $\chi_4(t,a)$ with $N = 1024$, $\tau_p = 10^2$, $\phi = 0.88$.
(f) Dynamical susceptibility $\chi_4(t,a)$ with $N = 1024$, $\tau_p = 10^{-2}$, $\phi = 1.055$.
}
\label{fig:dense}
\end{figure}

We use the velocity autocorrelation function $C_{\boldsymbol{v}\boldsymbol{v}}(t)$ to characterize time fluctuations.  Fig.~\ref{fig:vel}(d) shows that these decay over a timescale that depends strongly on the persistence time $\tau_p$ (see also SM \cite{suppmat} Fig.~S2), and weakly on $\phi$. Because particle motion is not hindered by crowding in the active liquid, large particle displacements easily occur along the spatially correlated velocity field, which maintains its structure over long timescales $\sim\tau_p \gg 1$. This is confirmed in Fig.~\ref{fig:vel}(e) which shows that the MSD is superdiffusive for times up to $\tau_p$, corresponding in real space to a spatially correlated displacement field [Fig.~\ref{fig:vel}(f)] which resembles the velocity field at $t=0$. The many vortices and abrupt changes of this velocity field, alongside its large-scale correlations, bear strong similarities with the family of active systems described as `turbulent'~\cite{wensink2012meso,nishiguchi_mesoscopic_2015,doostmohammadi2017onset,alert2022active} (see \texttt{SuppMovie1.mov} and \texttt{SuppMovie2.mov} \cite{suppmat}): remarkably, this occurs here in an active liquid without explicit aligning interactions.

On increasing density, the system becomes crowded, and diffusive motion sets in over a large timescale $\tau_\alpha (\phi) \propto D^{-1}$ that increases rapidly with $\phi$. At equilibrium, particle dynamics is triggered by infrequent thermally activated relaxation events characterized by a broad distribution of energy barriers, reflecting a rugged energy landscape~\cite{berthier2011theoretical}. This physical picture survives for modest values of the persistence times, as in Fig.~\ref{fig:arrest}(a), except that activated dynamics is now driven by a nonequilibrium colored noise, as recently studied in simpler active situations~\cite{woillez2019activated,woillez2020active}. Therefore, glassy dynamics for weak persistence qualitatively resembles passive systems~\cite{flenner2016nonequilibrium,berthier2019glassy}.

The physics is radically different when the persistence time is large, see Fig.~\ref{fig:dense}. Since particles are always in contact at these high densities, the natural time scale for relaxation is $\xi \sigma^2/\varepsilon = 1$.
This opens a time window, $1 \ll t \ll \tau_p$, where particle dynamics is nearly arrested and $| \boldsymbol{v}_i | \ll | \boldsymbol{p}_i |$, see Fig.~\ref{fig:arrest}(b).
The forces stemming from particle interactions then nearly balance the self-propulsion forces, and the system is close to mechanical equilibrium.  Such configurations correspond to local minima of an effective potential energy~\cite{mandal2020multiple,mandal2021how}, $U_{\rm eff} (\boldsymbol{r} ; \boldsymbol{p}) \equiv U(\boldsymbol{r}) - \sum_i \boldsymbol{p}_i \cdot \boldsymbol{r}_i$, which depends  on  both positions and propulsions. Because $U_{\rm eff}$ evolves slowly over a timescale $\tau_p \gg 1$, a mechanical equilibrium at time $t$ may be unstable at time $t+\tau_p$, because the propulsion forces will have significantly evolved via Eq.~(\ref{eq:p}). The loss of mechanical equilibrium triggers fast particle rearrangements~\cite{mandal2020extreme} (on a timescale $\sim1$), as the system relaxes towards a new minimum of $U_{\rm eff}$.  
This effect is illustrated in Fig.~\ref{fig:dense}(a), which shows the squared gradient of $U_{\rm eff}$ as a function of time. The succession of spikes correspond to rearrangement events. These results demonstrate the intermittent dynamics at large $\tau_p$, which differs qualitatively from the low-persistence case.

Intermittent dynamics between mechanical equilibria superficially resembles plasticity in slowly sheared amorphous solids~\cite{maloney2006amorphous}.
In this regime, the spatially correlated velocity field at $t=0$ does not dictate the displacements associated with structural relaxation.  To see this, note the weak correlations between Figs.~\ref{fig:dense}(c,d);
these snapshots should be contrasted with Figs.~\ref{fig:vel}(c,f) for the persistent liquid, where velocities and displacements are strongly correlated.  (The time scales were chosen such that the MSD is similar in both cases.)  In fact, the displacement field in the glassy case [Fig.~\ref{fig:dense}(d)] results from the accumulation of smaller particle rearrangements. This mechanism is distinct from the turbulent flows seen in the persistent liquid, and from activated relaxation events at weak persistence.

While it might be desirable to analyze relaxation events individually, $\tau_p$ is finite in our simulations, so changes in $U_{\rm eff}$ are not quasistatic. This makes individual rearrangements hard to characterize, a problem which is familiar from equilibrium supercooled liquids~\cite{berthier2011dynamical}.  Hence we analyze the dynamics using tools from that context, and start with the
non-Gaussian parameter \cite{kob1995testing,kob_dynamical_1997}
\begin{equation}
\alpha_2(t) =
\frac{ N^{-1} \sum_{i} \left<|\boldsymbol{r}_i(t) - \boldsymbol{r}_i(0)|^4\right> }{2(\Delta r^2(t))^2}   - 1,
\label{equ:alpha2}
\end{equation}
which is zero when the distribution of particle displacements is Gaussian. In the persistent liquid ($\phi\lesssim0.87$) the dominant sources of heterogeneity are the correlated velocities which decay on the time scale $\tau_p$.  Hence $\alpha_2(t)$ decays monotonically from the non-Gaussian velocity distribution at $t = 0$ to the Gaussian diffusive limit at $t \gg \tau_p$. For larger $\phi$, $\alpha_2(t)$ starts off similar to the active liquid,
but it increases again for $t>\tau_p$, leading to a maximum at a much longer timescale $\tau_\alpha \gg \tau_p$.
This peak shows that the slow dynamics are strongly non-Gaussian, supporting the picture of intermittent transitions \cite{chaudhuri_universal_2007} between minima of $U_{\rm eff}$.

To quantify the collective nature of the dynamics we study the dynamic susceptibility~\cite{berthier2011dynamical}
\begin{equation}
\chi_4(t, a) = N \left[  \langle Q^2(t,a) \rangle - \langle Q(t,a) \rangle^2  \right] ,
\label{equ:chi4}
\end{equation}
where $Q(t,a) = N^{-1} \sum_i \Theta(a - |\boldsymbol{r}_i(t) - \boldsymbol{r}_i(0)|)$ is a dynamic overlap and $a$ a length scale, see Fig.~\ref{fig:dense}(e). In dense passive systems, $\chi_4(t, a)$ is only sizeable if $t\sim \tau_\alpha$ and $a$ close to the typical cage size~\cite{lacevic2003spatially}, thus capturing the cooperative nature of activated relaxation. Here, $\chi_4(t, a)$ is large at all times up to $\tau_{\alpha}$, and is maximized for a given $t$ for $a \sim \sqrt{\Delta r^2(t)}$. These data quantitatively confirm that particle displacements are spatially correlated over a broad range of time scales in very persistent glassy systems. This contrasts with results at low persistence [Fig.~\ref{fig:dense}(f)], where the short-time correlations are absent, similar to the passive case.

We showed that strongly persistent systems support active liquid states with turbulent nonequilibrium flows and superdiffusive motion and that at high density, the dynamic arrest of these liquids is accompanied by complex spatio-temporal correlations, spanning a range of length and time scales, whose origin differs qualitatively from active glasses analysed so far. While slow dynamics are ubiquitous in crowded systems (both active and passive), a simple replacement of thermal noises by highly-persistent propulsions dramatically changes the mechanisms by which a liquid explores its energy landscape. The resulting dynamical processes are unusual and raise several questions for fiture work. Can the intermittent dynamics of the persistent active glass be understood by analysing the idealised limit of large $\tau_p$~\cite{mandal2021how}? How general are active turbulent states in systems without explicit aligning interactions~\cite{mandal2020extreme,kuroda2022anomalous}?  Can existing theories of velocity correlations be extended to such states~\cite{szamel2021long,marconi2021hydrodynamics}?

\begin{acknowledgments}
  We thank L. Cugliandolo, J. Tailleur, and F. van Wijland for discussions.
  This work was supported by a grant from the Simons Foundation (\#454935 L. Berthier).
\end{acknowledgments}

\bibliography{ref}

\end{document}